\documentstyle[aps,prl,epsf,twocolumn,epsfig,floats]{revtex}

\begin{document}

\draft


\wideabs{

\title{Superconducting Gap Anisotropy in Nd$_{1.85}$Ce$_{0.15}$CuO$_4$:  Results from 
Photoemission}
\author{N.P. Armitage, D.H. Lu, D.L. Feng, C. Kim, A. Damascelli, 
K.M. Shen, F. Ronning, and Z.-X. Shen}
\address{Department of Physics, Applied Physics and Stanford Synchrotron
        Radiation Lab., Stanford University, Stanford, CA 94305}
\author{Y. Onose, Y. Taguchi, and Y. Tokura}
\address{Department of Applied Physics, The
 University of Tokyo, Tokyo 113-8656, Japan}

\maketitle
\begin{abstract}
We have performed angle resolved photoelectron spectroscopy on the electron 
doped cuprate superconductor Nd$_{1.85}$Ce$_{0.15}$CuO$_4$.  A comparison 
of the leading edge midpoints between the superconducting and normal states 
reveals a small, but finite shift of 1.5-2 meV near the ($\pi$,0) position, 
but no observable shift along the zone diagonal near ($\pi$/2,$\pi$/2).  
This is interpreted as evidence for an anisotropic superconducting gap in 
the electron doped materials, which is consistent with the presence of 
$d$-wave superconducting order in this cuprate superconductor.
  
\end{abstract}


 } 


It is now generally agreed that the hole doped ($p$-type) cuprate 
superconductors have an order parameter whose majority component is of 
$d_{_x2-_y2}$ symmetry.  Angle resolved photoelectron spectroscopy (ARPES) 
has provided a key piece of early evidence by the demonstration of a large 
momentum space anisotropy of the electronic excitation gap\cite{Shen1}.  
Quantum interference devices have given a definitive attestation to the 
existence of an order parameter that changes sign in momentum 
space\cite{Kirtley}.  Corroborating evidence has been provided by a number 
of other techniques\cite{Scalapino}.

In contrast, the experimental and theoretical situation on the small family 
of electron-doped ($n$-type) cuprate superconductors\cite{Tokura}, such as 
Nd$_{1.85}$Ce$_{0.15}$CuO$_4$, has not reached a similar d$\acute{e}$tente.  
Tunneling\cite{Huang}, microwave penetration 
depth\cite{Wu}, and Raman \cite{Stadlober} measurements all yield evidence 
for some kind of nearly uniformly gapped Fermi surface.  In addition, 
Nd$_{1.85}$Ce$_{0.15}$CuO$_4$ is the only measured cuprate superconductor 
that does not exhibit the so-called zero-bias Andreev bound state on its 
(110) surface that has been interpreted as a consequence of a momentum 
space sign change in the superconducting order parameter\cite{Alff,Kashi}.
	  
However, recent work has called this picture of a uniformly gapped Fermi 
surface into some doubt.  There is evidence from measurements using 
scanning SQUID microscopy on tri-crystal films, that the electron-doped 
materials have an order parameter with a large $d$-wave 
component\cite{Tsuei}.  In addition, there has been speculation that the 
ordering at low temperature of the free 4$f$ moments on the rare earth Nd 
ions affects the usual microwave penetration depth measurements by altering 
the low temperature magnetic permeability\cite{Alff}.  A recent reanalysis 
of microwave data that incorporates a correction for the Nd spin ordering, 
as well as experiments on the related paramagnetic material 
Pr$_{2-x}$Ce$_{x}$CuO$_4$ have been interpreted as being more consistent 
with a $d$-wave scenario\cite{Cooper}.
	
On the theoretical side, there have been problems reconciling the $s$-wave 
experimental results with current models.  As in the hole doped case, 
strong on-site Coulomb repulsion in the copper oxygen plane strongly favors 
an anisotropic order parameter that has positive and negative lobes.
	 
We have recently restudied the Nd$_{1.85}$Ce$_{0.15}$CuO$_4$ system using 
high resolution ARPES and found two distinct features in the low energy 
spectra, which contrasts with the behavior of the p-types with hole doping, 
where the systematics of the single low energy feature interpolates between 
the prototypical undoped insulator and the optimally doped 
superconductor\cite{Armitage1}.  In the case of 
Nd$_{1.85}$Ce$_{0.15}$CuO$_4$, the spectra are dominated by a broad highly 
dispersive feature which resembles that found in the undoped insulator 
Ca$_2$CuO$_2$Cl$_2$\cite{Ronning}.  In addition, a weaker feature at lower 
binding energy ($<80$meV) is observed near the ($\pi, 0$) position.  These 
two features should not be viewed as deriving from seperate ``bands'' of 
independent origin, as they conspire together to give a Fermi surface with 
the expected Luttinger volume.
	 
In this letter we focus on an energy scale much lower than that of our 
previous work\cite{Armitage1} (below 5 meV).  The central intellectual 
issue is the anisotropy of the superconducting gap, which should manifest 
itself in the ARPES spectra as a displacement of the leading edge to higher 
binding energy as a gap opens up.
	
The single crystals of Nd$_{1.85}$Ce$_{0.15}$CuO$_4$ used in this study 
were grown by the traveling solvent floating zone method.  Details of this 
growth can be found elsewhere\cite{Onose}.

ARPES measurements were performed at beamline 5-4 of the Stanford 
Synchrotron Radiation Laboratory.  This system is a Scienta SES 200 
electron spectrometer coupled to a normal incidence monochromator (NIM) 
that is capable of better than 10 meV energy resolution and 0.2 degree 
angular resolution.  Data was taken in Scienta ``Angle'' mode, in which an 
entire $\pm$5.5$^{\circ}$ angular cut can be imaged simultaneously.  The 
incident photon beam angle was at approximately 45 degrees to the sample 
surface with its polarization in the direction of the angular cuts.  
Samples are positioned by a custom sample manipulator that can be cooled 
below 10 K.  The sample temperature was measured by a calibrated diode 
mounted near the measurement position.  The temperature difference between 
the diode position and the actual measurement position was found to be 
negligible previously from measurements on another diode that had been 
mounted to a sample holder in a identical fashion to the samples.
	
All Nd$_{1.85}$Ce$_{0.15}$CuO$_4$ measurements were taken with 16.5 eV 
photons, which has been found empirically to give a high photoionization 
cross-section.  Samples were aligned by Laue diffraction prior to vacuum 
chamber introduction.  Shiny flat surfaces resulted from cleaving the 
samples $in$ $situ$ by the top-post method at 10 K.  LEED was performed 
after the UPS measurments to check surface structure and quality.  For each 
measurement set, the temperature was cycled a number of times to ensure 
repeatability and an absence of problems from surface ageing.  No such 
problems were detected over the typical measurement time of 7 hours per 
k-space cut.  To improve statistics of these intrinsically low signal 
experiments the same temperature spectra were summed together.

     \begin{figure}[htb]
    \centerline{\epsfig{figure=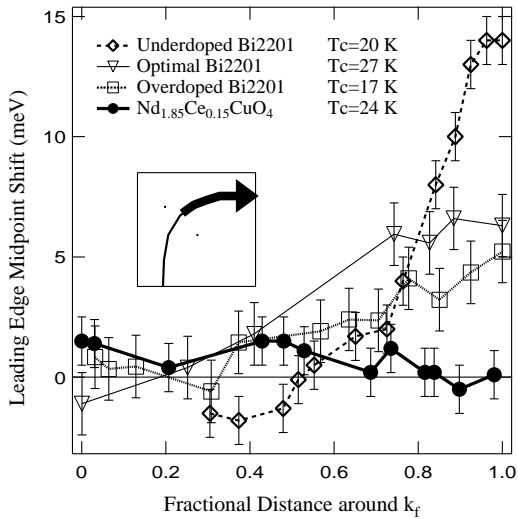,width=7.5cm}}
    \vspace{.2cm} 
    \caption{Plot of the LEM around the apparent Fermi 
     surface of a few cuprate superconductors of comparable ${T_c}$'s in units of the 
     fractional distance from  ($\pi/2, \pi/2$) to ($\pi, 0.3\pi$).}
    \end{figure}
     
In order to see signatures of superconductivity on a system with an 
expected small gap, it is crucial that one compare leading edge midpoints 
(LEMs) from the same sample above and below T$_c$.  In the past it has been 
standard practice to simply compare LEMs of the spectral weight onset to 
the Fermi energy of a reference gold sample, but this kind of analysis is 
not reliable in the small gap $n$-type systems.  To make this point more 
clear, in Fig.  1 we compare the trend of the LEMs of 
Nd$_{1.85}$Ce$_{0.15}$CuO$_4$ referenced to the Fermi energy to that of one 
plane bismuth based cuprates with similar ${T_c}$'s as our 
Nd$_{1.85}$Ce$_{0.15}$CuO$_4$ sample.  All spectra were taken at 
approximately 11 meV resolution and with the polarization parallel to the 
Cu-O bond.  The bismuth cuprate data is adapted from Ref.\cite{Bi2201}.  A 
few distinct trends are evident in the hole doped compounds.  There is a 
notable decrease in the maximum gap as one proceeds from underdoping to 
overdoping.  This is consistent with previous ARPES results as well as with 
other techniques.  There is a definite increase in the LEMs as one goes 
around the Fermi surface in accordance with one's d-wave expectation.  
Similar behavior is seen in YBa$_{2}$Cu$_{3}$O$_{7-x}$ and 
La$_{1.85}$Sr$_{0.15}$CuO$_4$.\cite{Schabel,Tep} In contrast, this analysis 
which reveals a large anisotropy on the hole doped compounds, does not 
reveal any clues as to anisotropy in Nd$_{1.85}$Ce$_{0.15}$CuO$_4$ as the 
LEM referenced to $E_F$ shows no systematic trends along its Fermi surface.  
This is a consequence of the fact that the superconducting gap by any 
measure is much smaller than in the hole doped materials and hence is 
obscured due to finite resolution and subtle changes in the lineshape 
around the Fermi surface.  If the systematics of the lineshape did not 
change very much around ${\bf k_F}$, one would still have a good measure of 
relative changes in the gap energy, despite not being able to measure its 
absolute magnitude exactly.  This has been done, for example, in the 
imaging of a gap in Bi2212 and YBa$_{2}$Cu$_{3}$O$_{7-x}$ that follows the 
$d$-wave functional form, despite large measured negative gap energies in 
the nodal region\cite{Schabel,Loeser}.  In the present case, due to the 
previously mentioned weaker component that appears near ($\pi, 0$) the 
systematics of the lineshape do change and a simple comparison with ${ 
E_F}$ is not valid.  One must then compare the shift of LEMs from spectra 
above and below Tc at the same momentum space point on the same sample in 
order to accurately measure the gap value.  This should remove extraneous 
effects of the final state and otherwise eliminate effects (e.g.  
photoionization matrix elements) other than a gap opening up.

We took coarse energy resolution data at 10K with over 500 energy 
distribution curves (EDCs) over a Brillouin zone octant and symmetrized 
across the zone diagonal.  Fig.  2 shows a 30 meV integration about the 
Fermi energy of these EDCs plotted over a Brillouin zone quadrant.  Along 
the circularly shaped Fermi surface, one sees three regions of high 
intensity.  The central one corresponds to the previously mentioned broad, 
highly dispersive feature moving into the integration window.  The other 
two are the intensity maxima of the smaller low binding energy feature.  In 
the inset is a typical low energy electron diffraction (LEED) pattern that 
shows bright well formed spots with a symmetry commensurate with that of 
the bulk.  The extreme surface sensitive nature of LEED is an excellent 
cross-check of the surface quality.  We took high resolution angle resolved 
cuts at two temperatures through ${\bf k_F}$ points in two different 
regions of momentum space as shown by the arrows in the figure ( ($0.46\pi, 
0.46\pi$) and ($\pi, 0.27\pi$) exactly, but hereafter refered to as near 
($\pi/2, \pi/2$) and ($\pi, 0.3\pi $) ).  In both cases the photon energy 
was 16.5 eV, polarization was parallel to the cut direction, and the energy 
resolution was $~$11 meV.

\begin{figure}[htb]
\centerline{\epsfig{figure=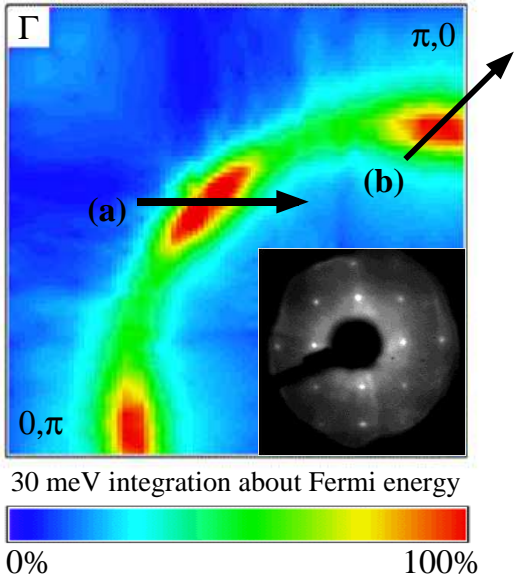,width=7cm}}
\vspace{.2cm} 
\caption{(color).  A 30 meV integration around the Fermi energy of each EDC plotted 
over the Brillouin zone quadrant.  Two high resolution angle resolved cuts 
[labeled (a) and (b)] were taken as marked by the arrows.  These are 
displayed in Fig.  3.  In the inset is the LEED pattern of the sample 
cleaved { \it in situ} at 10K. }
    \end{figure}

In Fig.  3a and 3b, we present color scale plots of the spectral function 
in a $\pm$5.5$^{\circ}$ ($\pm21\%$ $\pi$) angular windows from the cuts 
near ($\pi/2, \pi/2$) and ($\pi, 0.3\pi$) respectively.  In the spectra 
near ($\pi/2, \pi/2$) we find a dispersion which is universal for the 
cuprates in this region of the Brillouin zone.  A large broad feature 
disperses towards the Fermi energy, sharpens, and then disappears as it 
passes above E$_F$.  In 3b, which is a cut through ${\bf k_F}$ at 
approximately ($\pi, 0.3\pi$), the spectra are best characterized by a 
large hump feature that disperses only slightly while the smaller low 
energy feature disperses toward E$_F$, loses weight, and then disappears.  
This behavior has been characterized in more depth 
elsewhere\cite{Armitage1}.  In Fig.  3c and 3d, we compare spectra at 
0.5$^{\circ}$ steps taken at 30K (red) and 10K (blue) at a few near ${\bf 
k_F}$ positions for each cut.  In 3c from the spectra near ($\pi/2, \pi/2$) 
we see that, aside from some small thermal broadening, there is no 
temperature induced change.  This is quite different from that of spectra 
from the ${\bf k_F}$ crossing near ($\pi, 0.3\pi$) where there is a 
systematic displacement by $\sim$1.5-2meV of the leading edge to higher 
binding energy in the superconducting state.

\begin{figure}[htb]
\centerline{\epsfig{figure=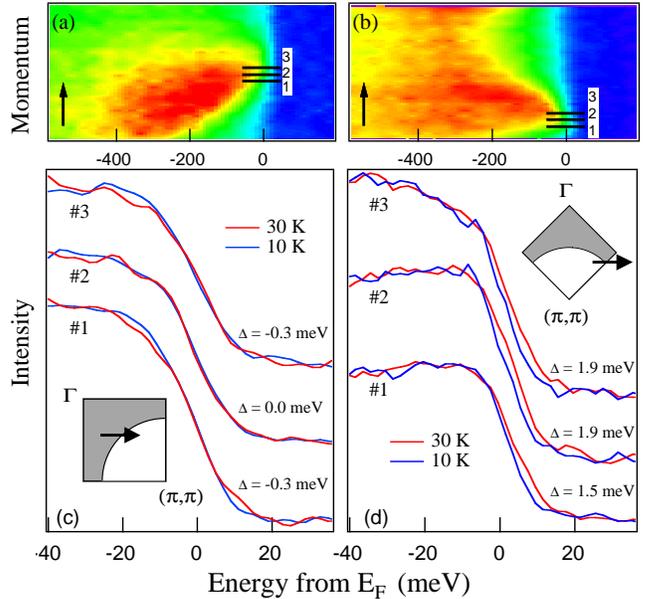,width=8.5cm}}
\vspace{.2cm} 
\caption{(color).  At the top labeled (a) and (b) are false color plots representing 
the intensity of the spectral function in a $\pm$5.5$^{\circ}$ angular 
windows ($\pm21\%$ $\pi$) near ($\pi/2, \pi/2$) and ($\pi, 0.3\pi$) 
respectively.  In (c) and (d) we compare EDCs taken at 30K (red) and 10K 
(blue) from the $k$-E regions represented by the horizontal bars in panels 
(a) and (b) respectively.  Arrows denote the direction that angular cuts 
are displayed as referenced to the arrows in insets of the Brillouin 
zone.  Gap values represent the results of the fitting procedures for 
the curves.}
\end{figure}
	   
In Fig.  4 there is an expanded image of a spectrum with one of the large 
temperature induced shifts in Fig.  3d.  To more systematically quantify 
the result, we fit the lineshape with a simple phenomenological model.  For 
broad spectral features at the Fermi level, the superconducting state 
spectra can be approximated by a 10K Fermi function multiplied by a linear 
spectral function whose onset edge is displaced from the Fermi energy by 
the superconducting gap energy.  Likewise, the normal state spectra can be 
modeled as that of a 30K Fermi distribution at finite temperature 
multiplied by the same (but non-displaced) linear spectral function.  These 
model spectra are convolved with a Gaussian of FWHM of 11 meV to simulate 
the experimental resolution.  The agreement between the experimental curves 
and the fits are quite good within this picture with a gap parameter of 1.9 
meV.
	    
\begin{figure}[htb] 
\centerline{\epsfig{figure=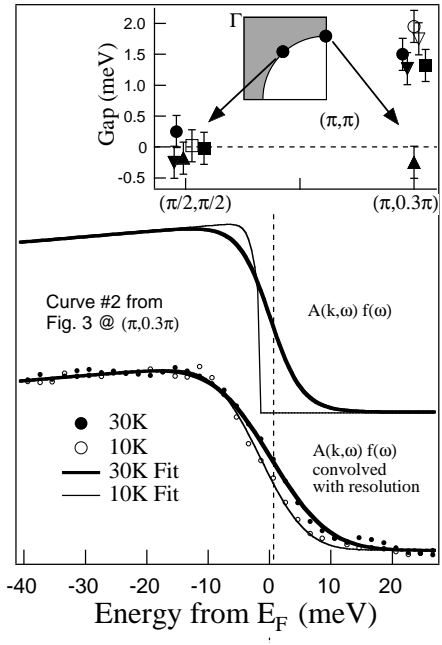,width=6cm}} 
\vspace{.2cm} \caption{ At the bottom of the panel are EDCs from ${\bf k_F}$ 
near ($\pi, 0.3\pi$) which is curve 2 in from Fig.  3d.  They are plotted along 
with a simple fit convolved with the resolution.  At the top is the same 
fit unconvolved with the experimental resolution.  The fitted gap parameter 
is 1.9 meV.  In the inset is a plot of the temperature dependence data from 
all 7 samples taken near ($\pi/2, \pi/2$) and ($\pi, 0.3\pi$).  The data are 
slightly offset from each other in the horizontal direction for display 
purposes.}
\end{figure}
	
The above temperature dependent measurements were repeated on 7 different 
samples (4 samples at both ${\bf k_F}$ positions, two only at ($\pi, 
0.3\pi)$ and one just at ($\pi/2, \pi/2$) ).  In the inset of Fig.  4 we 
plot the temperature dependent shift of the LEM at both ${\bf k_F}$ 
crossings for all samples.  Although there is some scatter in the data, one 
can see that, with the exception of one sample, they all show a 1.5-2 meV 
shift at the ($\pi, 0.3\pi)$ position and a negligible one at the ($\pi/2, 
\pi/2$) position.  This shift anisotropy is interpreted as a consequence of 
the opening of a superconducting gap that is maximal near ($\pi, 0)$ and 
minimal or zero along the zone diagonal.  This is consistent with the 
presence of a d$_{_x2-_y2}$ superconducting order parameter.  However, we 
cannot rule out a very anisotropic $s$-wave order with a small gap along 
the zone diagonal that is below our detection limit (approximately the same 
as the scatter in the data at ($\pi/2, \pi/2$), $<0.3meV$).  The reason for 
the lack of a temperature dependent shift on the one sample is unknown.  
Its spectral features and dispersions were consistent with the others.  It 
is possible that it suffered from being in poor thermal contact with the 
cold stage.

The 2$\Delta/k_BT_c$ ratio in the narrow doping range of superconducting 
n-type materials is typically measured to be smaller than those in the 
hole-doped ones.  There is no consensus on where they should fit 
analogously into the hole doped phase diagram, but perhaps the T$^2$ 
dependence of their resistivity suggests that they are more overdoped-like.  
We may then expect a low 2$\Delta/k_BT_c$ ratio.  The gap value of 1.5-2 
meV obtained with our technique is consistent with, but slightly smaller 
than, the gap values reported by other 
techniques\cite{Huang,Wu,Stadlober}.  There are a few possibilities 
for this.  These other measurements typically take place at 4.2 K (0.15 
T$_c$), in contrast to our measurement at approximately 0.5 T$_c$.  At our 
intermediate temperatures the superconducting gap may not have opened 
fully.  In addition, there may be some background contribution that 
partially obscures the gap signal.  Lastly, it is well documented that 
photoemission analysis based on LEMs consistently underestimates the 
maximum gap value (in some cases by as much as a factor of 2) as compared 
to the intrinsic value defined as the quasiparticle energy at the antinodal 
position.\cite{Loeser}.
	
In conclusion, we believe there is evidence for a anisotropic 
superconducting gap in the electron doped material that is consistent with 
a $d_{_x2-_y2}$ pairing state.  Despite the strong differences between the 
$p$ and $n$-type compounds in the larger scale electronic 
structure($\sim$1eV), their superconductivity appears to share the same 
symmetry and is therefore possibly of similar origin.  The experimental 
data was recorded at the Stanford Synchrotron Radiation Laboratory which is 
operated by the DOE Office of Basic Energy Science, Division of Chemical 
Sciences and Material Sciences.  Additional support comes from the Office 
of Naval Research: ONR Grants N00014-95-1-0760/N00014-98-1-0195.  The 
crystal growth work was supported in part by Grant-in-Aids for Scientific 
Research from the Ministry of Education, Science, Sports, and Culture, 
Japan, and the New Energy and Industrial Technology Development 
Organization of Japan (NEDO).



\end{document}